# Fine-Tuning, Complexity, and Life in the Multiverse*

**Mario Livio**
Dept. of Physics and Astronomy, University of Nevada, Las Vegas, NV 89154, USA
E-mail: dr.mario.livio@gmail.com
and
**Martin J. Rees**
Institute of Astronomy, University of Cambridge, Cambridge CB3 0HA, UK
E-mail: mjr36@cam.ac.uk

**Abstract**
The physical processes that determine the properties of our everyday world, and of the wider cosmos, are determined by some key numbers: the 'constants' of micro-physics and the parameters that describe the expanding universe in which we have emerged. We identify various steps in the emergence of stars, planets and life that are dependent on these fundamental numbers, and explore how these steps might have been changed — or completely prevented — if the numbers were different. We then outline some cosmological models where physical reality is vastly more extensive than the 'universe' that astronomers observe (perhaps even involving many 'big bangs') — which could perhaps encompass domains governed by different physics. Although the concept of a *multiverse* is still speculative, we argue that attempts to determine whether it exists constitute a genuinely scientific endeavor. If we indeed inhabit a multiverse, then we may have to accept that there can be no explanation other than anthropic reasoning for some features our world.

___________________

*Chapter for the book Consolidation of Fine Tuning

# 1 Introduction

At their fundamental level, phenomena in our universe can be described by certain laws—the so-called "laws of nature" — and by the values of some three dozen parameters (e.g., [1]). Those parameters specify such physical quantities as the coupling constants of the weak and strong interactions in the Standard Model of particle physics, and the dark energy density, the baryon mass per photon, and the spatial curvature in cosmology.

What actually determines the values of those parameters, however, is an open question. Many physicists believe that some comprehensive "theory of everything" yields mathematical formulae that determine all these parameters uniquely. But growing numbers of researchers are beginning to suspect that at least some parameters are in fact random variables, possibly taking different values in different members of a huge ensemble of universes — a multiverse (see e.g., [2] for a review). Those in the latter 'camp' take the view that the question "Do other universes exist?" is a genuine scientific one. Moreover, it is one that may be answered within a few decades. We address such arguments later in this chapter, but first we address the evidence for 'fine tuning' of key parameters.

A careful inspection of the values of the different parameters has led to the suggestion that at least a few of those constants of nature must be fine-tuned if life is to emerge. That is, relatively small changes in their values would have resulted in a universe in which there would be a blockage in one of the stages in emergent complexity that lead from a 'big bang' to atoms, stars, planets, biospheres, and eventually intelligent life (e.g., [3–6]).

We can easily imagine laws that weren't all that different from the ones that actually prevail, but which would have led to a rather boring universe — laws which led to a universe containing dark matter and no atoms; laws where you perhaps had hydrogen atoms but nothing more complicated, and therefore no chemistry (and no nuclear energy to keep the stars shining); laws where there was no gravity, or a universe where gravity was so strong that it crushed everything; or the cosmic lifetime was so short that there was no time for evolution; or the expansion was too fast to allow gravity to pull stars and galaxies together.

Some physicists regard such apparent fine-tunings as nothing more than statistical flukes. They would claim that we shouldn't be surprised that nature seems 'tuned' to allow intelligent life to evolve — we wouldn't exist otherwise. This attitude has been countered by John Leslie, who gives a nice metaphor. Suppose you were up before a firing squad. A dozen bullets are fired at you, but they all miss. Had that not happened, you wouldn't be alive to ponder the matter. But your survival is still a surprise — one that it's natural to feel perplexed about.

Other physicists are motivated by this perplexity to explore whether 'fine tuning' can be better understood in the context of parallel universe models. In this connection it's important to stress that such models are consequences of several much-studies physical theories — for instance cosmological inflation, and string theory. The models were not developed simply to remove perplexity about fine tuning.

Before we explore some prerequisites for complexity, it is instructive to examine a pedagogical diagram that demonstrates in a simple way the properties of a vast range of objects in our universe. This diagram (Figure 1), adapted from Carr & Rees [7], shows the mass vs. size (on a logarithmic scale) of structures from the subatomic to the cosmic scale. Black holes, for example, lie on a line of slope 1 in this logM–logR plot. A black hole the size of a proton has a mass of some $10^{38}$ protons, which simply reflects how weak the force of gravity is. Solid objects such as rocks or asteroids, which have roughly the atomic density, lie along a line of slope 3, as do animals and people. Self-gravity is so weak that its effects are unnoticeable up to objects even the size of most asteroids. From there on, however, gravity becomes crucial, causing for instance planets to be spherical, and by the time objects reach a mass of about 0.08 $M_\odot$, they are sufficiently squeezed by gravity to ignite nuclear reactions at their centers and become stars. The bottom left corner of Figure 1 is occupied by the subatomic quantum regime. On the extreme left is the 'Planck length' — the size of a black hole whose Compton wavelength is equal to its Schwarzschild radius. Classical general relativity cannot be applied on scales smaller than this (and indeed may break down under less extreme conditions). We then need a quantum theory of gravity. In the absence of such a theory, we cannot understand the universe's very beginnings (i.e., what happened at eras comparable with the Planck time of $10^{-43}$ seconds).

Despite this unmet challenge, it's impressive how much progress has been made in cosmology. In the early 1960s, there was no consensus that our Universe had expanded from a dense beginning. But we have now mapped out, at least in outline, the evolution of our universe, and the emergence of complexity within it, from about a nanosecond after the Big Bang. At that time, our observable universe was roughly the size of the solar system, and characterized by energies of the order of those currently realized at the Large Hadron Collider (LHC) near Geneva. Nucleosynthesis of the light elements gives us compelling corroboration of the hot and dense conditions in the first few seconds of the universe's existence (see, e.g., [8] for a recent review).

The cosmic microwave background (CMB) provides us not only with an astonishingly accurate proof for a black-body radiation state that existed when the universe was 400,000 years old, but also with a detailed map of the fluctuations in temperature (and density), $\Delta T/T \sim 10^{-5}$, from which eventually structure emerged. Peaks in the power spectrum of the CMB fluctuations, mapped with great accuracy by the WMAP and Planck satellites, can, even without any other information, offer precise determinations of a few cosmological parameters (e.g., [9, 10]), such as the fractions of baryonic matter, dark matter, and so-called 'dark energy' in the cosmic energy budget (Figure 2).

The latter is a mysterious form of energy latent in empty space which has a negative pressure, and causes the cosmic expansion to accelerate. It was discovered through observations of Type Ia supernovae [11, 12]. Since then, however, its existence has been confirmed through other lines of evidence, including the CMB, Baryon Acoustic Oscillations, and the Integrated Sachs Wolfe effect (see [13] for a brief review). The simplest hypothesis is that the dark energy has the same properties as the cosmological constant 'lambda' which Einstein introduced in his original equations, but it is possible that it has more complicated properties. In particular, it could change with time, and could correspond to just one of many possible vacua. In addition, many lines of evidence have led to the realization that some form of gravitating dark matter outweighs ordinary baryonic matter by about a factor of five in galaxies and clusters of galaxies. Here are four: (i) flat rotation curves in galaxies extending out beyond the stellar disk; (ii) the motions of galaxies in clusters of galaxies; (iii) the temperature of the hot gas in clusters of galaxies; (iv) gravitational lensing. All of these measure the depth of the gravitational potential well in galaxies or clusters and reveal the presence of mass that does not emit or absorb light. While all the attempts to detect the constituent particles of dark matter have so far been unsuccessful (see e.g., [14] for a review), this may not be so surprising when we realize that there are some ten orders of magnitude between the currently observed mass-energies and the GUT unification energy, where these particles could hide. Moreover, there are other options such as axions, or ultra-low-mass bosons.

Dark matter provided the scaffolding onto which the large-scale structure formed. In fact, while some uncertainties about the details remain (see, e.g., Chapter by Debora Sijacki in this volume), computer simulations can generally reproduce the types of structures we observe on the galactic and cluster scale while starting from the fluctuations observed by Planck and WMAP (see e.g., [15]).

Similarly, a combination of hydrodynamics, thermodynamics and nuclear physics has led to a fairly satisfactory understanding of the main processes involved in stellar structure, star formation, evolution, and stellar deaths (e.g., [16, 17]), as well as the formation of planetary systems. Thanks to observations in the past two decades (especially by the Kepler Space Observatory), we now know that the Milky Way contains about one Earth-size habitable-zone planet for every six M dwarfs [18], which makes the prospects of finding extrasolar life (at least in simple form) with planned or proposed telescopes more promising [19–21].

Given our current understanding of the evolution of our universe, and of galaxies, stars and planets within it, we may attempt to identify what the prerequisites for life are. Since our knowledge of the processes involved in the emergence of life, however, lags far behind our comprehension of fundamental physical processes, we shall only list those very basic requirements that we think should apply to any generic form of complexity.

## 2 Prerequisites for Complexity

There are (at least) five prerequisites for the emergence of complexity in a universe; these prerequisites would not be fulfilled in a counterfactual universe where the fundamental constants were too different from their actual values.

'Counterfactual' exercises of this type are useful for developing an intuition about the role of physical constants in the evolution of the universe and in the emergence of complexity. Similar studies are used by historians to explore various "what if?" scenarios, such as speculating what might have happened had Archduke Franz Ferdinand of Austria not been shot by a Serb nationalist in Sarajevo in 1914. Biologists similarly wonder about how the history of life on Earth might have changed had the dinosaurs not been wiped out by an asteroid impact.

If the acceptable range of values for some parameter is small, we would define it as 'fine tuned.' We shall briefly discuss the extent to which this is the case for some key parameters.

## 2.1 Constraints on Gravity

As numerical simulations of structure formation in the universe have demonstrated, gravity enhances density fluctuations. In our universe, gravity caused the denser regions to lag behind the cosmic expansion and to form the sponge-like structure that characterizes the universe on its largest scales. Eventually, gravity led to the formation of galaxies at the density peaks, of stars, and of planets. Stellar evolution also represents one continuous battle with gravity, the latter pushing the stellar central densities and temperatures to higher and higher values. On the surface of planets gravity played crucial roles in keeping an atmosphere bound and in bringing different elements into contact to initiate the chemical reactions that eventually led to life. But gravity in our universe is a very weak force — the ratio of the repulsive electric force between two protons to their gravitational mutual attraction is $e^2/Gm^2_p \sim 10^{36}$. The reason gravity becomes important on the scale of large asteroids and higher, is that large objects have a net electric charge that is close to zero, so gravity wins once sufficiently many atoms are packed together.

Figure 1 allows us to make a first attempt to examine what would happen in a universe in which the values of some "constants of nature" are different. How would Figure 1 be different if gravity were not so weak? The general structure of the diagram would remain the same, but there would be fewer powers of ten between the subatomic and the cosmic scales. Stars, which effectively are gravitationally bound nuclear fusion reactors, would be smaller in such a universe and would have shorter lives. If gravity were much stronger, then even small solid bodies (such as rocks) might be gravitationally crushed. If gravity's strength were such that it would still have allowed tiny planets to exist, life forms the size of humans would be crushed on the planetary surface. Overall, the universe would be much smaller and there would be less time for complexity to emerge. In other words, to have what we may call an "interesting" universe (in the sense of complexity), we must have many powers of ten between the microscale and the cosmic scale, and this requires gravity to be very weak. It is important to note, however, that gravity does not need to be fine-tuned for complexity to emerge. In fact, a universe in which gravity is ten times weaker than in our universe, may be even more "interesting" in that it would allow bigger stars and planes, and more time for life to emerge and evolve.

## 2.2 CP Violation — More Matter than Antimatter

The Big Bang in our universe created a slight excess (by about one part in three billion) of matter over antimatter. It has been shown that for such an imbalance to be created, baryon number and CP symmetry (charge conjugation and parity) had to be violated in the Big Bang, as well as interactions being out of thermal equilibrium (the so-called 'Sakharov conditions' [22]). Had the matter–antimatter imbalance not existed, particles and antiparticles would have all annihilated to form only radiation (what we observe today as the CMB), leaving no atoms and therefore no galaxies, no stars, no planets and no life. Within the Standard Model of particle physics the most promising source of CP violation appears to be in the lepton sector, where it generates matter-antimatter asymmetry via a process known as leptogenesis. If, however, CP violation in the lepton sector will be experimentally determined to be too small to explain the matter-antimatter imbalance (as was the case with the Cabibbo-Kobayashi-Maskawa matrix in the quark sector: [23]), physics beyond the Standard Model would be required.

## 2.3 Fluctuations

'Curvature fluctuations' were imprinted into the Universe at a very early era. Their amplitude is almost independent of scale. Many theorists suspect that they originated as quantum fluctuations during an inflationary phase, when the presently observable universe was of microscopic size. The physics of this ultra-early era is of course still speculative and uncertain. However we know from observations that the fluctuations gave rise to temperature fluctuations that grew to $\Delta T/T \sim 10^{-5}$ at the time of recombination.

These fluctuations were crucial for the emergence of complexity. If the early universe had been entirely smooth, then even with the same microphysics the universe today would have been filled only with cold hydrogen and helium. Stars, galaxies, and indeed people would never have formed. The parameter that measures how 'rough' the universe is, is called Q. At recombination the temperature fluctuations across the sky $\sim\Delta T/T$ are of order Q. There is no firm theoretical argument that explains why it has the observed value of about $10^{-5}$ (see, e.g., [1, 24] for a discussion). Computer simulations have offered a huge boost to the credibility of our current ΛCDM model by showing that under the action of gravity and gas dynamics, the fluctuations observed in the CMB would evolve into galaxies with the morphological properties and luminosity functions observed, grouped into clusters whose statistical properties also match the observations.

But what would happen in a counterfactual universe where Q were different from its actual value, but all other cosmic parameters stayed the same? If the amplitude of the fluctuations were larger, say $Q \sim 10^{-4}$, masses of about $10^{14}$ M$_\odot$ would condense at a cosmic age of about 300 million years. At that time, Compton cooling on the (then warmer) microwave background would allow the gas to collapse into huge disk galaxies. The virial velocity in large-scale systems scales as $Q^{1/2}c$, and these giant galaxies would find themselves (after some $10^{10}$ yrs) in clusters with masses of $\gtrsim 10^{16}$ M$_\odot$. A universe with $Q \sim 10^{-4}$ would have an even larger range of non-linear scales than ours. It would offer more spectacular cosmic vistas; and the only reason why it might be somewhat less propitious for life is that stars in the galaxies would be more close-packed, rendering it less likely that a planetary system could remain undisrupted by a passing star for long enough to permit biological evolution. However, if Q were even larger ($Q \gtrsim 10^{-3}$), conditions would be very unfavorable for life. Enormous masses of gas (far larger than a cluster of galaxies in our present universe) would condense out early on, probably collapsing to massive black holes — an environment too violent for life.

[Incidentally, any observers who could exist in a high-Q universe would find it far more challenging to interpret and quantify their surroundings. Because Q is small in our actual universes, even the largest non-linear structures are very small compared to the cosmic horizon (they are smaller by a factor of order $Q^{1/2}$). We can therefore observe a large number of independent patches, and define average smoothed-out properties of the universe — the mean density, etc. — and use the standard homogeneous cosmological models as a good approximation. By analogy, a sailor watching ocean waves can meaningfully describe their statistical properties, because even the longest wavelength is small compared to the distance of the horizon. In contrast, an astronomer in a high-Q universe would resemble a climber in a mountain landscape, where one peak could dominate the view, and averages aren't well-defined].

What about the other extreme, a 'smoother' universe with $Q \lesssim 10^{-6}$? In this case the disruptive dark energy would push protogalaxies apart before they had a chance to collapse. Even if the dark energy weren't there, any galaxies that formed in a lower-Q universe would be small and rather loosely bound (and forming later than in our actual universe). At $Q \gtrsim 10^{-6}$ stars would still form, but material enriched in heavy elements, and ejected via stellar winds or supernovae, may escape from the shallow gravitational potential wells, not allowing for second-generation stars and planetary systems to form. For values of Q that are significantly smaller than $10^{-6}$, there would be inefficient radiative cooling and therefore stars would not form within a Hubble time. The conclusion from this discussion (summarized also in [6], and see Figure 3), is that for a universe to be conducive for complexity and life, the amplitude of the fluctuations should best be between $10^{-6}$ and $10^{-4}$, and therefore not particularly finely tuned.

## 2.4 Non-Trivial Chemistry

For life to emerge, the universe requires nuclear fusion. Fusion not only powers the stars, but nucleosynthesis at the hot stellar centers also forges elements such as carbon, oxygen, iron and phosphorus, all of which are essential for life as we know it. In general, many of the elements in the Periodic Table participate in the complex chemistry required for the formation of planets and the evolution of their biospheres.

To obtain the nuclear fusion reactions that lead to the creation of the Periodic Table requires a certain balance between the strength of the electromagnetic force (that repels two protons from each other) and the strong nuclear force (that attracts them). This balance, in our universe, where the strong nuclear force is about a hundred times stronger than the electromagnetic force, is responsible for the fact that we don't have atomic numbers higher than 118. Had the ratio of the two interactions been much smaller, carbon and heavier elements could not have formed, but the necessary tuning is not excessive.

Similarly, much has been written about Fred Hoyle's prediction of the existence of a 7.65 MeV resonant level of $^{12}C$ [25, 26]. However, while the prediction itself was indeed remarkable, the degree of fine-tuning required for the energy of that level is not fantastic (e.g., [27, 28], and see [29] for a recent study of this energy level).

The topic of chemistry actually allows us to examine a much more extreme counterfactual universe — a 'Nuclear-Free Universe' — in which hydrogen is the only element that exists. Surprisingly, on the large scale such a universe would not look much different from ours. Gravity would ensure that galaxies would still form, and even stars would shine (albeit generally for shorter times) by releasing their gravitational energy as they contract to form white dwarfs and black holes. Even Jupiter-like planets composed of solid hydrogen could exist. Of course no complexity or life of the types we are familiar with will emerge in such a universe (only perhaps something similar to Fred Hoyle's science fiction concept of *The Black Cloud* [30]).

## 2.5 'Tuned' Cosmic Expansion Rate

The results from the Planck Satellite depicted in Figure 2 (in combination with observations of Baryon Acoustic Oscillations, lensing reconstruction and a prior on the Hubble constant) give for the cosmic energy budget $\Omega_m \sim 0.3$, $\Omega_\Lambda \sim 0.7$, with baryons making less than 5% of this budget [9]. If the cosmic acceleration is indeed driven by a cosmological constant (energy of the physical vacuum, with an equation of state parameter $w = P/\rho = -1$), then the acceleration will continue forever. It is clear, however, that if the dark energy density would have dominated over the matter density (dark matter + baryons) much earlier in the life of our universe, galaxies would never have formed (this is also dependent on the value of Q, see discussion in the next section). This means that for complexity to arise, some constraints are needed on the ratios of $\Omega_m/\Omega_\Lambda$ and $\Omega_b/\Omega_{DM}$ (where $\Omega_b$ denotes the baryon fraction and $\Omega_{DM}$ the dark matter fraction). The second ratio is crucial because even though dark matter dominates over baryonic matter in our universe, without the latter there would be no stars, no planets, and no life.

As an aside we should note that the nature of the dark energy that propels the cosmic acceleration is one of the most fascinating puzzles in modern cosmology (and one that may not be solved until we have a better understanding of the granular structure of space-time on the Planck scale). Despite its importance for fundamental physics, the dark energy hardly affected any astrophysical phenomena in our universe; in contrast, the evolution of our universe so far — the emergence of and morphology of galaxies, clusters, and so forth — has been dominated by the effects of dark matter.

# 3 The Multiverse

As far as we can tell, the laws of physics and the values of the cosmological parameters are the same throughout our entire observable universe. But the observable universe is limited by the horizon, which is determined by the finite age of our universe. What lies beyond this Hubble volume? The homogeneity and isotropy of our observable universe, with the absence of any perceptible gradient across it (to the $10^{-5}$ level) suggest (though of course do not prove) that the same laws continue to apply thousands of times further. Indeed, many arguments suggest that galaxies beyond the horizon outnumber those we see by a vast factor — perhaps so vast that all combinatorial options would occur repeatedly, and we'd all, far beyond the horizon, have avatars.

Furthermore, some models for the inflationary phase lead to what has been dubbed 'eternal inflation' [31, 32]. According to these models, our Big Bang could be just one 'pocket universe' in a huge ensemble — one island of space-time in a vast archipelago. This scenario also fits well with the 'landscape' concept of string theory, in which there are some $10^{500}$ meta-stable vacua solutions, of which our universe is but one [33, 34]. So the question arises: How large is physical reality? [1]

The first thing to realize is that because we live in an accelerating universe, galaxies are disappearing over an 'event horizon,' so we will not observe their far future (rather as we can't observe the fate of an object that falls into a black hole after it has crossed the horizon). If the acceleration continued, then after about a trillion years observers in the remnant of the Milky Way (or its merged product with the Local Group) would not be able to see (again, even in principle) any galaxy other than their own. This does not mean that those galaxies whose light would have been stretched beyond the cosmic scale would not exist.

Moreover, galaxies that are already beyond our current horizon will never become observable, even in principle. Yet most researchers would be relaxed about claims that these galaxies exist, in the same way that in the middle of the ocean you expect that an ocean extends beyond the terrestrial horizon. These never-observable galaxies would have emerged from the same Big Bang as we did. But suppose that we imagine separate Big Bangs. Are space-times completely disjoint from ours any less real than regions forever unobservable which are the aftermath of 'our' Big Bang? Surely not — so these other universes too should count as parts of physical reality.

Similarly, while we cannot observe any free quarks, we believe that quarks exist, because the Standard Model of particle physics has successfully passed many experimental tests. Likewise, we are disposed to believe in what Einstein's theory tells us about the metric within black holes (inside the event horizon), because General Relativity has gained high credibility by being tested in numerous observations and experiments.

If we can develop a theory that makes numerous predictions that are testable (and are confirmed) in the observable part of the universe, then we should be prepared to accept its predictions in unobservable parts.

We currently have no theories of microphysics that are 'battle tested' above the energies reachable in the biggest particle accelerators. These energies are exceeded throughout the first nano-second after the Big Bang. Theorists who model the inflation era therefore make assumptions about the physics. Some such assumptions predict eternal inflation; others do not. Some predict the landscape scenario; others do not. The details of this physics are already somewhat constrained (by, for instance, the observed properties of the fluctuations in the CMB), but we are still far from being able to prove or disprove a model like Linde's 'eternal inflation' [32]. We should therefore be open-minded about how far the aftermath of 'our' Big Bang extends beyond our horizon, and also about whether other Big Bangs exist as part of physical reality. Once we are willing to entertain the notion that a multiverse may exist, an even more intriguing question arises: Are the laws of physics and the values of the physical constants the same in other members of this ensemble of universes or are they different?

---

[1] It's perhaps necessary, especially in addressing philosophical readers, to inject a clarification at the start. Many would define 'the universe' as 'everything there is' — and if that's the definition, then there plainly cannot be more than one. If there are other domains (perhaps originating in other big bangs, and perhaps differing from our observable domain in size, content, or dimensionality) then we should really define the whole enlarged ensemble as 'the universe.' We then need a new word — 'metagalaxy' for instance — to denote the domain to which cosmologists and astronomers have observational access. However, so long as this whole idea remains speculative, it is probably best to leave the term 'universe' undisturbed, with its traditional connotations, even though this then demands a new term, 'multiverse,' for the whole (still hypothetical) ensemble.

If they are different, than what we call 'laws of nature' may be no more than local bylaws governing just our cosmic patch. Moreover, many of these pocket universes could be still- born or sterile. That is, the physical laws prevailing in them (or the values of the parameters) may preclude the emergence of any kind of complexity, and life in particular. They simply may not satisfy one or more of the prerequisites we discussed in Section 2.

The mere possibility that physical reality can encompass such a multiverse provides a strong motivation to develop the lines of thought outlined in Section 2 — to explore a variety of counterfactual universes, with different values of physical constants, to ascertain which ranges of parameters would allow complexity. The identification and selection of such 'biophilic' universes constitutes what has been dubbed anthropic reasoning (e.g., [4, 6, 36–38]). Obviously we humans find ourselves not in a typical member of the multiverse, but in a typical domain in the subset of universes that allows complexity and life to emerge and evolve [38]. Copernican humility can only be taken so far. We live on an ordinary terrestrial planet orbiting an ordinary star in its habitable zone. The observable universe may contain as many as two trillion galaxies [39], and our universe may be only one member of an ensemble of some $10^{500}$ universes. But our universe is not 'typical.'

To give a simple analogy (which we believe was first suggested by physicist Leonard Susskind), suppose you wake up in the morning and think: "What am I?" It seems that a natural answer may be "I am an insect," since insects have the largest biomass of terrestrial animals. It is estimated that at any time there are some $10^{19}$ insects alive. The reason that this argument is false is that by being able to wonder "what am I?" we have already selected a small subset of the animal kingdom. On the other hand, we can argue that the probability that the answer to "what am I?" is "I am Leonardo da Vinci" is still very small.

To be able to actually determine the ranges of all the parameters that allow complexity to develop and the probability for its emergence is currently beyond what physics can achieve, since it requires a knowledge of all the probability distributions and the correlations among them. What we can currently do is a 'poor man's simplified version of this daunting task — going beyond the discussion of Section 2, where parameters were varied one at a time, and analyzing the 'anthropic' domain in a two parameter diagram. For instance, we can depict different values of the dark energy (assumed to be due to a cosmological constant $\Lambda$), and the amplitude of the fluctuations Q (Figure 3; [6]). Structures can only form so long as gravity overwhelms the repulsive effect of $\Lambda$. A higher value of Q implies earlier formation of structure and therefore higher values of $\Lambda$ would still be anthropically allowed in such a universe.

Another two-parameter example (see [1]) takes Q and the density of dark matter as two parameters that could vary. If the dark matter density were higher than in our actual universe, the cosmic expansion would become matter-dominated at an earlier stage, allowing more time for the growth of structures from initial fluctuations. So the anthropically allowed values of Q would then extend downward. (This contrasts with the effect of higher values of Λ, which extend the allowable Q upwards).

We are currently far from having any theory that determines the values of Λ or Q or the dark matter density (and we know even less about the relative likelihood of various combinations of these constants or how they might be correlated). Still less do we have a cosmological model that can put a 'measure' on the probability density of various combinations. But if we did, we would then have another way of testing — and in principle refuting — whether the 'fine tuning' was due to anthropic selection. We could do this by examining whether we existed in a 'typical' part of the anthropically allowed multiverse, or whether the tuning was even more 'special' than anthropic constraints required. This line of reasoning can be illustrated by a simple analogy:

Even if we knew nothing about how stars and planets formed, we would not be surprised to find that our Earth's orbit was fairly close to circular: had it been highly eccentric, water would boil when the Earth was at perihelion, and freeze at aphelion — a harsh environment unconducive to our emergence. However, a modest orbital eccentricity, up to 0.1 or so, is plainly not incompatible with life. But suppose it had turned out that the Earth moved in a near-perfect circle with an eccentricity of 0.000001. Some quite different explanation would then be needed: anthropic selection from orbits whose eccentricities had a 'Bayesian prior' that was uniform in the range 0–1 could plausibly account for an eccentricity of 0.1, but not for one as tiny as this.

The methodology requires us to decide what range of values is compatible with our emergence. It also requires a specific theory that gives the relative Bayesian priors for any particular value within that range. With this information, one can then ask if our actual universe is 'typical' of the subset in which we could have emerged. If it is a grossly atypical member even of this subset (not merely of the entire multiverse), then we would need to abandon our hypothesis that the numbers were anthropically selected.

Most physicists would consider the 'natural' value of 'dark energy' in the 'landscape' to be large, because it is a consequence of a very complicated microstructure of space. Current evidence suggests that the 'dark energy' has an actual value 5–10 times below the anthropically allowed maximum (other parameters being constrained to their actual values). That would put our universe between the 10th or 20th percentile of universes in which galaxies could form. In other words, our universe isn't significantly more special, with respect to Λ, than our emergence demanded. But suppose that (contrary to current indications), observations showed that Λ made no discernible contribution to the expansion rate, and was thousands of times below the threshold, not just 5–10 times. This 'overkill precision' would (like the precisely circular orbit in the analogy given earlier), raise doubts about the hypothesis that Λ was equally likely to have any value, and suggest that it was zero for some fundamental reason (or that it had a discrete set of possible values, and all the others were well about the threshold).

In principle we could, when theoretical models were more advanced — analyze other important parameters of physics in the same way, to test whether our universe is typical of the habitable subset that could harbor complex life. The methodology requires us to decide what values are compatible with our emergence. It also requires a specific theory that gives the probability of any particular value. For instance, in the case of Λ, is there a set of discrete vacuua or a continuum of values? In the latter case we need to know whether all values are equally probable, or whether the probability density is clustered at a low value.

## 4 Conceptual Shifts

The introduction of the multiverse and of anthropic reasoning has generated considerable controversy, sometimes even accompanied by passionately negative reactions from a number of physicists. We have already discussed the first main objection — the sentiment that envisaging causally disconnected, unobservable universes is in conflict with the traditional "scientific method." We have emphasized that modern physics already contains many unobservable domains (e.g., free quarks, interiors of black holes, galaxies beyond the particle horizon). If we had a theory that applied to the ultra-early universe, but gained credibility because it explained, for instance, some features of the microphysical world (the strength of the fundamental forces, the masses of neutrinos, and so forth) we should take seriously its predictions about 'our' Big Bang and the possibility of others.

We are far from having such a theory, but the huge advances already made should give us optimism about new insights in the next few decades. Indeed, even at this early stage, eternal inflation and the landscape scenario already make some predictions that are in principle testable. For example, in eternal inflation, our universe is expected to have a very small (~$10^{-4}$) negative curvature (a "bubble"). Therefore, future measurements of spatial curvature (including measurements of the 21 cm transition) could falsify eternal inflation (e.g., [40]). Similarly, accelerator experiments can (in principle) generate conditions in which a number of metastable vacuum solutions are possible, thereby testing the premises of the landscape scenario. It is also possible (although the probability for such an event is very low), for another inflating bubble to pop close and collide with our bubble universe, leaving an imprint in our CMB (e.g., [41]). These simple examples demonstrate that even though the multiverse idea is still in its infancy, this scenario constitutes a bona fide topic (albeit speculative) of scientific discourse, rather than metaphysics.

We have also pointed out that an anthropic explanation can be refuted, if the actual parameter values are far more 'special' than anthropic constraints require.

Many physicists still hope that a unique, self-consistent theory of the universe will unambiguously determine the values of all the physical parameters. This is a lofty goal, but the history of science has already demonstrated that a quest for first-principle explanations for everything can fail. The great astronomer Johannes Kepler tried to find answers to two questions: (i) Why there were precisely six planets (only six were known at his time), and (ii) What it was that determined the spacings among planetary orbits. Eventually he thought he found the answer, and he published it in his book *Mysterium Cosmographicum* (originally published in 1597; [42]). Kepler's answer was at the same time impressive and absolutely wrong. He constructed a model of the solar system in which the five Platonic solids were embedded one inside the other and together in a surrounding sphere. This created exactly six spaces (like the number of planets), and by choosing the order of the solids in a particular way, the spacing agreed with the relative sizes of the orbits to within 10 percent. The model was impressive because Kepler used mathematics to explain observed phenomena. It was completely wrong because Kepler didn't understand at the time that neither the number of the planets nor their orbits were fundamental phenomena that required first-principles explanations. Rather, we understand today that both the number of planets and their orbits are accidental variables whose values are determined by the environmental conditions in which the planetary system formed. Earth's orbit is special only insofar as it is in the habitable zone around the Sun.

The same may be true for at least some of the parameters of our universe, such as the values of $\Lambda$ and $Q$. These may be random variables in the multiverse, whose only "explanations" are offered by anthropic arguments. In view of our current ignorance as to what is truly fundamental and what is not, we should keep an open mind to all options.

More specifically, some 'constants' may be truly universal and others not. As an analogy (which we owe to Paul Davies) consider the form of snowflakes. Their ubiquitous six-fold symmetry is a direct consequence of the shape of water molecules. But snowflakes display an immense variety of patterns because each is molded by its micro-environments: how each flake grows is sensitive to the fortuitous temperature and humidity changes during its growth as it falls towards the ground. If physicists achieved a fundamental theory, it would tell us which aspects of nature were direct consequences of the bedrock theory (just as the symmetrical template of snowflakes is due to the basic structure of a water molecule) and which (like the distinctive pattern of a particular snowflake) were contingencies, taking many values across the multiverse.

If we indeed live in a multiverse, this would be a fifth (and in some sense the grandest) Copernican revolution. First, Copernicus showed that we are not at the center of the solar system; Harlow Shapley showed that the solar system is not at the center of our galaxy; the Kepler Space Observatory showed that there are billions of planetary systems in the Milky Way; Edwin Hubble and his namesake telescope have shown that there are trillions of galaxies in the observable universe; and now we realize that our observable domain may be only a tiny part of an unimaginably large and diverse ensemble. The next few decades will hopefully shed some light on the reality of the multiverse.

One thing, however, is clear. Our cosmic horizons have expanded precisely as fast as human knowledge. Every one of the five Copernican revolutions listed above marked an incredible human achievement. In that sense, we remain of central significance to our universe.

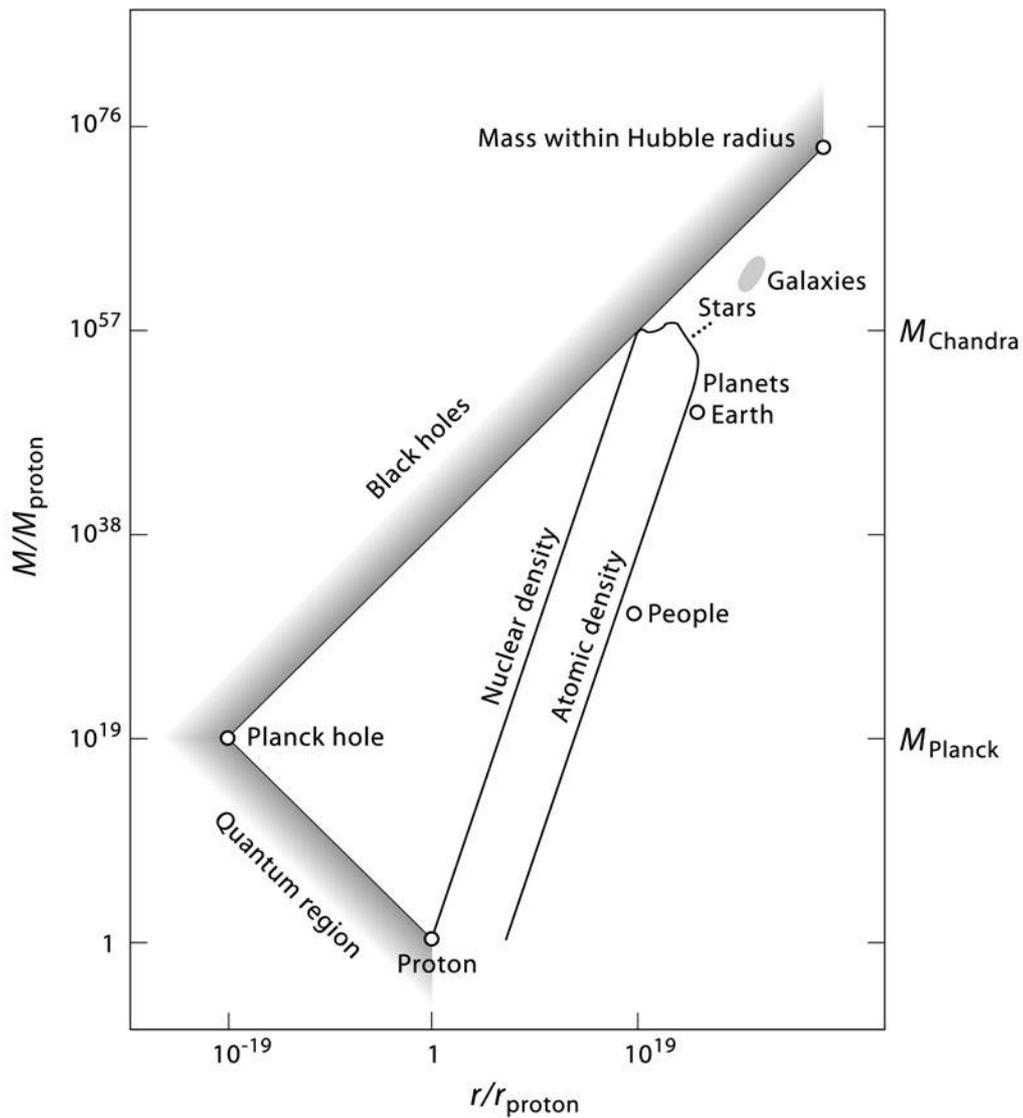

Figure 1. This diagram summarizes the scales of stars, planets, black holes and other bodies in a log-log plot of mass against radius. Ordinary lumps lie on the line of slope 3. The mass, in units of the proton mass, scales roughly as the cube of the radius. That line would eventually cross the black hole line (of slope one) at a mass of about 100 million suns. However it is curtailed before it can do so. The reason is that for any mass above about that of Jupiter (containing more than $10^{54}$ atoms) would be crushed by gravity to a higher density than an ordinary solid. If G were different, the shape of the diagram would not change much, but the number of powers of 10 between the scale of stars and of atoms would scale as the inverse 3/2 power.

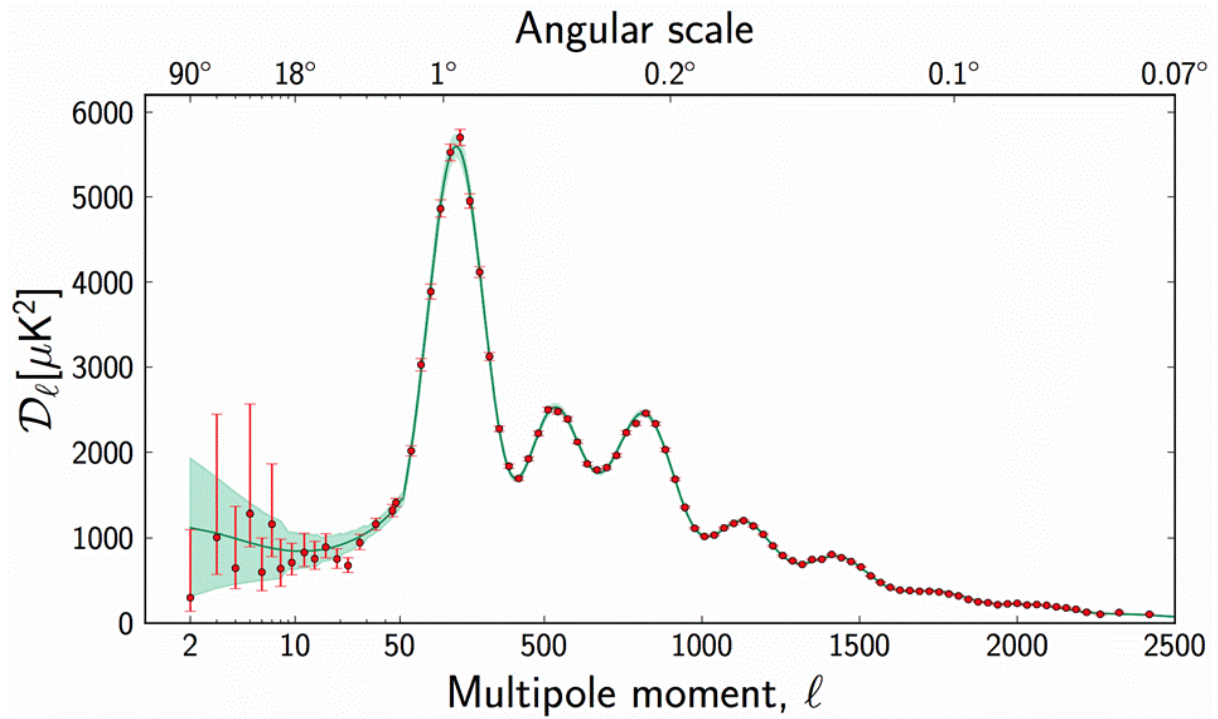

Figure 2. The fluctuations in the microwave background on different angular scales. The data come from the Planck spacecraft. The angular scale of the strongest peak is consistent with a 'flat' universe and the relative heights of the other peaks determine the baryon and dark matter densities.

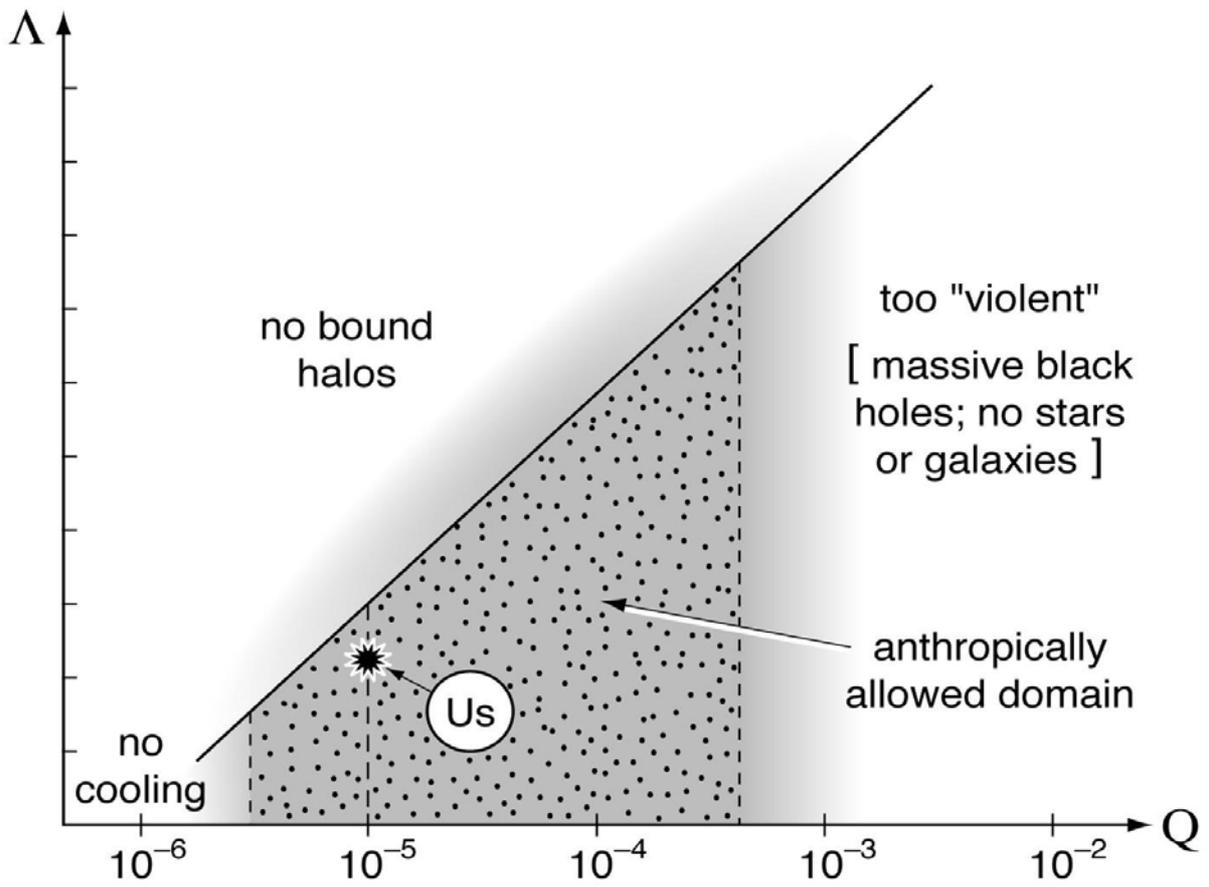

Figure 3. Plot of the cosmological constant Λ versus amplitude of fluctuations in cosmic microwave background Q. Shaded-dotted region shows conditions that allow for the existence of complexity.